\documentclass[useAMS,usenatbib]{mn2e}
\usepackage{times}
\usepackage{graphicx}
\usepackage{epstopdf}

\title[J08069+1527: A new hybrid sdBV.]{J08069+1527: A newly discovered high amplitude, hybrid subdwarf  B pulsator}
\author[A. S. Baran et al.]{A. S. Baran,$^{1,2}$\thanks{E-mail:
asb@iastate.edu}\thanks{ASB and JTG: Visiting Astronomers, Kitt Peak National Observatory, National Optical Astronomy Observatory, which is operated by the Association of Universities for Research in Astronomy (AURA) under cooperative agreement with the National Science Foundation} 
J. T. Gilker,$^{3}$ M. D. Reed,$^{3}$ R. H. {\O}stensen,$^{4}$ J. H. Telting,$^{5}$ K. Smolders$^{4}$ \newauthor L. Hicks$^{3}$ and R. Oreiro$^{4,6}$\\
$^{1}$ Mt. Suhora Observatory, Cracow Pedagogical University, ul. Podchorazych 2, 30-084 Krakow, Poland\\
$^{2}$ Iowa State University, 12 Physics Hall, Ames, IA 50011, USA\\
$^{3}$ Department of Physics, Astronomy, and Materials Science, Missouri State University, Springfield, MO 65897 USA\\
$^{4}$ Instituut voor Sterrenkunde, K.U.~Leuven, Celestijnenlaan 200D, 3001 Leuven, Belgium\\
$^{5}$ Nordic Optical Telescope, 38700 Santa Cruz de La Palma, Spain\\
$^{6}$ Instituto de Astrof\'isica de Andaluc\'ia, Glorieta de la Astronom\'ia, s/n 18008 Granada, Spain\\
}
\begin{document}

\date{}

\pagerange{\pageref{firstpage}--\pageref{lastpage}} \pubyear{2010}

\maketitle

\label{firstpage}

\begin{abstract}
We present our discovery of a new hybrid pulsating subdwarf B star, J08069+1527. The effective temperature and surface gravity of 28,500$\pm$400\,K and 5.37$\pm$0.04\,dex, respectively, place this object inside the instability strip and also among other pulsating hot subdwarfs of a hybrid nature, right next to another fascinating star: Balloon\,090100001. From this proximity, we anticipated this star could pulsate in both high and low frequency modes. Indeed, our analysis of photometric data confirmed our prediction. We detected two peaks in the high frequency region and two other peaks at low frequencies. In addition, the amplitude of the dominant mode is very high and comparable to the dominant peaks in other hybrid subdwarf B stars. Since this star is bright, we performed time-series low resolution spectroscopy. Despite a low signal-to-noise (S/N) ratio, we were able to detect the main peak from these data. All our results strongly indicate that J08069+1527 is a high amplitude pulsating hot subdwarf B star of hybrid nature. By analogy to the other pulsating sdB star, we judge that the dominant mode we detected here has radial nature. Future stellar modeling should provide us with quite good constrains as p- and g-modes presented in this star are driven in different parts of its interior.

\end{abstract}

\begin{keywords}
stars: subdwarfs, asteroseismology.
\end{keywords}

\section{Introduction}
Hot subdwarf B (sdB) stars are horizontal branch stars with masses about 0.5 M$_\odot$ and very thin, in mass, hydrogen envelopes. Their average effective temperature and surface gravity are about 30\,000~K and $\log g \sim$\,5.5, respectively. Although the future evolution of the sdB stars to the white dwarf cooling track is generally accepted, their formation prior to the sdB stage is under debate resulting in several mechanisms that involve single-star or binary evolution \citep[e.g.][]{dcruz96,han02,han03}.

The detection of pulsations in sdB stars opened a way to study their interiors and may help to understand their evolution prior to the horizontal branch. First, short period oscillations were found by \cite{kilkenny97} in EC\,14026-2647 (now officially named V391\,Hya). According to theoretical models, these pulsations are attributed to pressure modes and are driven in the outer part of the star \citep{charp97}. Several years later, \cite{green03} announced another kind of variability in sdB stars. In this case, however, the pulsations were identified with gravity modes. Following \cite{fontaine03} they are driven deeper in the sdB stars than the pressure ones. Short-period sdBV (sdB Variable) stars are located at higher, while long-period sdBV stars at lower effective temperatures.

The pulsation modes of the two kinds of sdBVs probe different regions, and stars displaying both types are particularly interesting since their interiors can be better constrained. Such hybrid pulsations were first found in DW \,Lyn \citep{schuh06} and another three objects with clear pulsations at both short and long periods have so far been recognized; Balloon\,090100001 \citep{baran05, oreiro05}, V391\,Peg \citep{lutz09} and RAT\,0455+1305 \citep{baran10}. Another star, CS\,1246, has similar $\log T_{\rm eff}$ and $\log g$ and one high amplitude short periodicity detected. Unfortunately, because of noisy data, the detection threshold in the low frequency region is too high. The best data are in the r' filter, and the amplitude spectrum calculated from these data contains some excess signal in the low frequency region, but more data are needed to confirm if CS\,1246 is a hybrid pulsator. This should encourage others to obtain more observations of CS\,1246. Extremely high precision photometry obtained with the Kepler spacecraft has revealed several long period sdBVs \citep{reed10, baran11} and one short period sdBV \citep{kawaler10} that appear to also show hybrid modes, at least intermittently, that are so weak that they can only be detected using satellite data. An unusual hybrid pulsator in an eclipsing binary, 2M1938+4603, was found to show an exceptionally large number of weak ($<$0.5\,mma) pulsation modes spanning all frequencies from the long period to the short period regime \citep{ostensen10a}.

%The Kepler satellite has also recently discovered several candidate hybrid sdBV stars \citep{reed10,kawaler10,ostensen10a,baran11} from its survey phase with follow-up data to be obtained over the next several years.

\begin{figure}
\includegraphics[width=83mm]{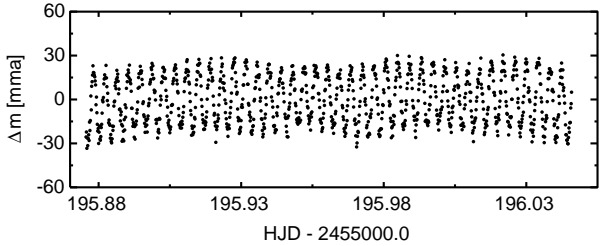}
\caption{Light curve in a BG40 filter obtained in December 2009 at KPNO.}
\label{kpnodata}
\end{figure}

Typical amplitudes of sdBV stars showing high frequency oscillations are below 1\%, which corresponds to 10\,mma. Most hybrid stars have unusually high amplitudes (few per cent), making them extremely desirable. Since sdBV stars tend to be faint ($12>m_{\rm V}>17$\,mag) high amplitude pulsators are easier to observe, even with 1\,m class telescopes.

As was shown by \cite{baran10}, for hybrid stars the dominant modes in both frequency domains appear at very similar frequencies. These are $\sim$ 0.3\,mHz for gravity modes and $\sim$ 2.8\,mHz for pressure modes. As for gravity modes, it may not be so surprising as the peaks are detected in a very small range, but for the pressure modes it is more astonishing. This may indicate that these stars are alike in internal structure and/or evolutionary status on the horizontal branch. In this paper we present our discovery of a new hybrid sdBV star with a high amplitude short period mode. We obtained data on four nights, including the discovery one.

%Among many modes, they have one high frequency mode with amplitude of about a few to several per cent and their highest amplitude mode in low frequency region is also higher than typical for sdBVs with amplitudes sometimes reaching 0.5\%. There are, however, sdBVr stars with high amplitude modes not showing, down to the detection threshold, any low frequency modes \citep{kilkenny99}. On the other side, their effective temperatures and surface gravities differ significantly (by 4\,kK and 0.4\,dex on average) from those of hybrid stars which are clustered around 29,5\,kK and 5.35\,dex.

%To date, we know of four hybrid sdBV stars detected from ground. The list of them is given in \cite{baran09}. There is also another star in the hybrid region in $\log T_{\rm eff}$ and $\log g$ plane with one high amplitude mode detected. It is CS\,1246 \citep{barlow10}. Unfortunately, data on CS\,1246 are too noisy which causes a high detection threshold in the low frequency region. The lowest noise is in r' filter and the amplitude spectrum calculated from this data contains some signal excess in the low frequency region, but more data are needed to confirm its possibly intrinsic nature. This should encourage others to obtain more observations of CS\,1246.

\begin{figure}
\includegraphics[width=83mm]{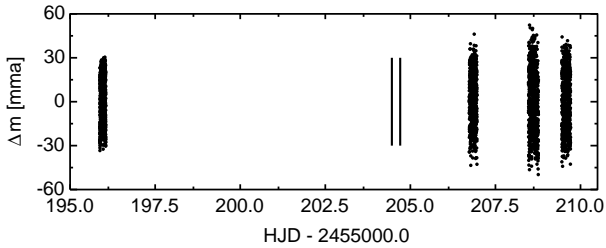}
\caption{All data obtained on J08069+1527. The first night was taken at KPNO, while the second at Baker and the last two at Mercator. Note that data from the the first two nights were taken in a BG40 filter while during the last two we used a B filter (Geneva system). Each set of points stands for one night. Two vertical lines indicate  coverage of spectroscopic data.}
\label{alldata}
\end{figure}

%As was shown by \cite{baran09}, for hybrid stars the dominant modes in both frequency domains appear at very similar frequencies. These are $\sim$ 0.3\,mHz for gravity modes and $\sim$ 2.8\,mHz for pressure modes. As for gravity modes, it may not be so surprising as the frequency range is very small, but for the pressure modes it is more astonishing. This may indicate that these stars are alike in an internal structure or evolutionary status on the horizontal branch. These objects are very convenient from a technical reason. Since all sdBV stars are relatively faint, stars with high amplitude modes are extremely desirable as detection of those modes is very easy, even with small telescopes. 

%The discovery data, taken at KPNO with the 2.1\,m telescope were also sufficient to detect a signal in the low frequency domain so this star seems to be another hybrid pulsator.

\section{J080656.7+1527}
The target J08069+1527, has UV magnitudes FUV\,=\,13.68, NUV\,=\,13.86 in the GALEX All-sky Imaging Survey \citep[AIS]{martin05} observation obtained on February 2, 2006. After comparing the UV magnitudes with visual photometry in the NOMAD survey ($B$\,=\,13.88, $V$\,=\,14.23, $R$\,=\,14.83), we flagged the target as a likely sdB star, and it entered a list of bright spectroscopy targets that served as a poor weather backup for a photometric observation run in the NOT search for pulsating sdBs in January 2009 \citep{ostensen10b}. The extracted spectrum was fitted to a model atmosphere spectrum, using the same procedure as for the other sdBs observed during the search program. The derived effective temperature of 28,877($\pm$202)\,K and surface gravity of 5.34($\pm$0.03)\,dex along with its encouraging brightness made it a priority target for the next observing season. Recently, \cite{vennes10} have published their first results from a survey of UV-excess stars in the GALEX archive, and J08069+1527 is included in their sample. Their determination of T$_{\rm eff}$ and $\log g$ is fully consistent with ours, within the errors.

\begin{table}
 \centering
 \begin{minipage}{83mm}
  \caption{Observational Log.}
  \label{log}
  \begin{tabular}{@{}cccccc@{}}
  \hline
   Date [UT]      &   hours   &   exposure   &   filter           &   site          & observer \\
 \hline
30 Dec 2009  &     4.0      &         12s       &  BG40          & KPNO       & ASB,JTG \\
10 Jan 2010   &     5.4      &         30s       &  BG40          & Baker       &  LH,MDR  \\
11 Jan 2010   &     7.0      &         31s       &  B(Geneva) & Mercator  & KS \\
12 Jan 2010   &     6.2      &         31s       &  B(Geneva) & Mercator  & KS \\
\hline
20 Jan 2009   &      ----     &        600s       & spectrum     & NOT         & RO \\
07 Jan 2010   &      6.6     &          30s       & spectra        & NOT          & JHT \\
\hline
\end{tabular}
\end{minipage}
\end{table}

%\begin{figure}
%\includegraphics[width=83mm]{fig1.eps}
%\caption{Location of hybrid stars in logT$_{\rm eff}$ and $\log g$ diagram. All known hybrid stars are marked with colors: Balloon\,090100001-blue, DW\,Lyn-red, V391\,Peg-green. Location of RAT\,0455 is marked with RAT.}
%\label{hr}
%\end{figure}

\section{Time series photometry}
\subsection{Discovery data}
We performed time-series photometry of J08069+1527 on 30 December 2009 using the KPNO 2.1\,m telescope. We used an Apogee CCD with a BG40 filter (wide band covering UBV filters range) with 12\,sec exposure times. Seeing conditions allowed us to set 2$\times$2 binning resulting in very short readout times of about 1\,sec. To determine if a star is variable or not, we used near-real-time software (developed at Missouri State University) during our observations to monitor the light curve as it was obtained. Right after we started to collect data, it was obvious that this star changes its brightness and the amplitude of this variability is relatively high. In fact, the amplitude was so big that even from our real-time software it was easy to estimate the period and amplitude of the variability. We estimated less than 10\,minutes for the period and about 20\,mmag for the amplitude. Based on its physical parameters, we assumed this variability to be caused by pulsation and realized that this is another (out of about seven known so far) high amplitude pulsating sdB star and decided to carry on observations on that night as long as possible. In total we collected 4 hours of data.

\subsection{Supplemental data}
After we found J08069+1527 to be an sdB pulsator with a high amplitude mode, we tried to obtain more time-series photometry to extend time coverage to lower noise level in the amplitude spectrum and increase frequency resolution. Unluckily, another opportunity to observe this star appeared only after 10 days. This long gap makes it harder to combine all data together since a window function gets too complicated and so does pre-whitening. Follow-up data were collected at Baker Observatory and with the Mercator telescope on La Palma. At Baker Observatory we used a Photometrics RS-1340 CCD with a BG40 filter. Exposure times were set to 30\,sec along with 1\,sec readout time. At the Mercator telescope we used the Merope CCD camera, which was recently upgraded with a new E2V frame transfer CCD with 2048$\times$3074 illuminated pixels \citep{ostensen10c}. Here we used a B filter (Geneva system) setting the data cadence period to about 31\,sec. We include all this information in Tab.\ref{log}.

%\begin{figure}
%\includegraphics[width=83mm]{fig3.eps}
%\caption{Window function of data taken at KPNO shifted to the frequency of the main mode.}
%\label{kpno-window}
%\end{figure}

We calibrated all star images for instrumental effects (bias, dark, flat field) and extracted brightnesses of target and field stars. In the next step we calculated differential photometry using a few comparison stars. All these tasks were done by means of the {\em Daophot} package \citep{stetson87} with a graphical user interface ({\em FinRed}, developed by ASB). Because the field of view only covers a few-arcminutes on the sky, the first order extinction (differential extinction) was negligible and long term variations (on the scale of run duration) were de-trended by subtracting a cubic spline curve. We assume here that any change in brightness, on time scales of about 6\,hours or longer, is not intrinsic to the star and is likely caused by non-perfectly removed extinction, either first or second order or other atmospheric effects.

In the next step, treating all nights separately, we transformed all brightnesses to fluxes and subtracted a mean value. Resulting light curves were then centered around zero. Next, we analyzed these data by means of a Fourier analysis. We present our discovery light curve of J08069+1527 obtained at KPNO on 30\,Dec\,2009 in Fig.\ref{kpnodata}. One can note that besides short period variability some long term variability is also present in the light curve. We present all data in Fig.\ref{alldata}.

\begin{figure}
\includegraphics[width=83mm]{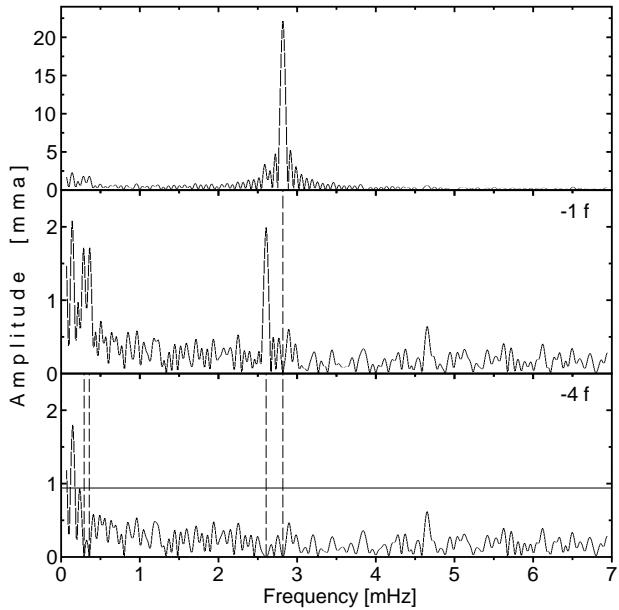}
\caption{Amplitude spectrum of KPNO data. The top panel shows the original data while the subsequent panels show pre-whitening as indicated by dashed lines. The solid horizontal line at 0.937\,mma represents S/N\,=\,4. The significant peak left in the amplitude spectrum appears at 2\,hrs (half of the run length) and requires more data to be confirmed.}
\label{kpno-ft}
\end{figure}

\subsection{Amplitude spectrum}
Although we have data taken with three different telescopes, we decided not to combine them all together. It is because data from KPNO and Baker, taken in the same filter, are 10\,days apart. When combining such data, we arrived with a higher noise level in the amplitude spectrum and a very complex window function. In the case of data from Baker and Mercator, although collected on consecutive nights, they are not in the same filter. As a result, during the pre-whitening process, we can obtain residual (fake) peaks caused by different amplitudes and perhaps phases of modes observed in different filters.

Since the telescope at KPNO is the biggest among those we used, data collected on that site are characterized by the smallest noise level in their amplitude spectrum. These data allow us to detect the highest number of peaks at S/N$>$4 used as our threshold. We detected two peaks in the high frequency region and two in the low frequency region. The dominant peak has a high amplitude and a frequency similar to some other sdBV stars discovered so far. \cite{baran10} showed that all hybrid sdBV stars detected from ground have a main dominant high frequency peak at similar frequencies around 2.8\,mHz. The main peak in J08069+1527 is also located in the same region. What is more, its effective temperature and surface gravity is almost the same as other hybrids, particularly Balloon\,090100001. In addition, we found two peaks in the low frequency region. Precise estimation of their significance is rather hard since not all the signal can be pre-whitened from that frequency region. This increases the real noise level. A detection threshold used in our analysis is an average through all frequencies in the amplitude spectrum, which might be a source of underestimation in the low frequency region. To estimate the noise level in that region more reliably, we used the adjoining 0.35--1\,mHz region and assumed it to be valid for those two peaks. Although the detection threshold increased, they still remained significant. Amplitude spectra with all pre-whitening steps calculated from KPNO data are shown in Fig.\ref{kpno-ft}.

\begin{figure}
\includegraphics[width=83mm]{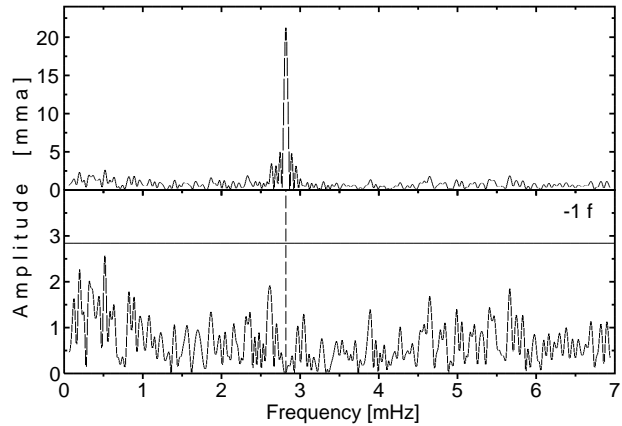}
\caption{Amplitude spectrum and the residual one calculated from data taken at Baker. The solid horizontal line at 2.84\,mma represents S/N\,=\,4.}
\label{baker-ft}
\end{figure}

In pre-whitening data taken at Baker, we could safely secure only one peak at the same frequency (within errors) as the dominant one from analysis of KPNO data. Others, if present are below the detection threshold. Only one peak detected here should not be surprising as the average noise level reaches 0.7\,mma so only peaks higher than 2.8\,mma could be detected with sufficient confidence. Looking at the result from KPNO, none but the dominant peak satisfies this condition. An amplitude spectrum as well as the residual one is shown in Fig.\ref{baker-ft}

\begin{figure}
\includegraphics[width=83mm]{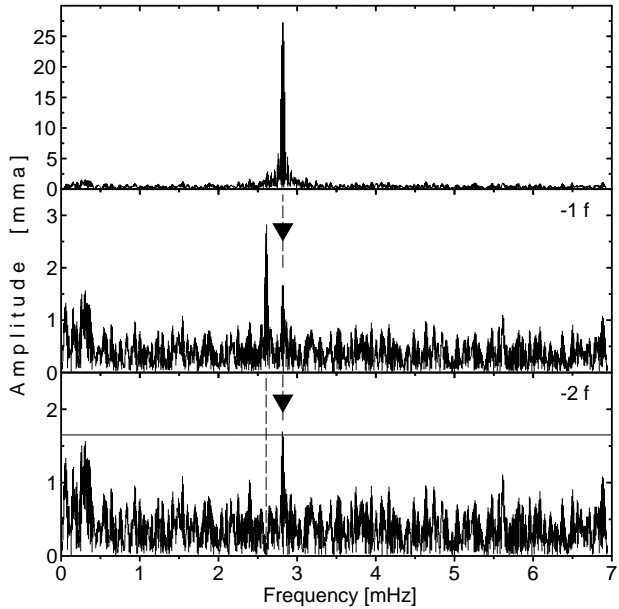}
\caption{Amplitude spectrum with pre-whitening steps calculated from data taken at Mercator. The solid horizontal at 1.65\,mma line represents S/N\,=\,4.}
\label{mercator-ft}
\end{figure}

At Mercator we obtained data on two consecutive nights. Although here we have better resolution and more points collected, we could not detect more peaks compared to the KPNO data. There is some signal excess in the low frequency domain. Unfortunately, the noise level is too high to confirm any peaks in this region. We can only confirm two peaks in the high frequency region. The dominant peak we detected here has the same frequency as the one detected in KPNO and Baker data while the other has a similar frequency but barely exceeds the error limits. However, apart from amplitude errors, frequency errors are sensitive to the length of observations. That is why we may consider the frequencies derived from Mercator data as the more precise ones. On the other hand, after removing the dominant peak, the residual amplitude spectrum still contains a small peak at the frequency of the removed one (Fig.\ref{mercator-ft}). It can mean that this peak is physically unstable, or there may be a mode beating, or some instrumental effects that cause either frequency, or amplitude to be different on each night. From Tab.\ref{list} we can see that frequencies are not changing within the given errors. Comparing amplitudes is harder since at Mercator we used a different filter. By comparing KPNO and Baker data, which were obtained in the same filter we can conclude that the amplitude has not changed over 10\,days (or is changing with periodicity of about 10\,days). Trying to solve this issue we chopped Mercator data into nightly chunks and Fourier analyzed them separately. From the Fourier solution, we found that timing issues between nights is not bigger than 0.5\,sec. However, the amplitudes from these two nights are substantially different. We derived 25.72(42)\,mma and 29.12(42)\,mma on the first and second nights, respectively.  Since the residual of the pre-whitened data roughly equals the difference of the amplitudes from the two nights, we can conclude that the cause is the change in amplitude.

\begin{table*}
 \centering
% \begin{minipage}{83mm}
  \caption{Results of the pre-whitening process of all data treated separately by site. The phases are given at
mean epoch (HJD) of each night: 2455195.960754, 2455206.844458 and 2455209.065208 for KPNO, Baker and Mercator, respectively. The numbers in parentheses are the errors of the last digits. Gaps in the cells mean that no signal was detected either in Baker or Mercator data at frequencies detected in KPNO data.}
  \label{list}
  \begin{tabular}{@{}cccccccccc@{}}
  \hline
                          & \multicolumn{3}{c}{KPNO} & \multicolumn{3}{c}{Baker} & \multicolumn{3}{c}{Mercator} \\ 
mode                & f [mHz]               & A [mma]       & phase [rad] & f [mHz]         & A [mma]     & phase [rad]  & f [mHz]           & A [mma]      & phase [rad] \\
\hline
f$_{\rm A}$      &     0.2926(47)    &     1.68(18)  & 3.74(12)      &                      &                     &                       &                         &                     & \\
f$_{\rm B}$      &     0.3577(53)    &     1.63(17)  & 1.63(12)      &                      &                     &                       &                         &                     & \\
f$_{\rm 1}$      &     2.81862(28)  &  22.06(16)  & 2.237(7)       &  2.8188(7)  &  21.23(52)  & 3.46(02)      &   2.81858(4)  & 27.29(30)  & 2.46(1) \\
f$_{\rm 2}$      &     2.6059(31)    &     1.98(16)  & 4.45(8)      &                      &                      &                      &   2.6119(4)     &   2.83(30)  & 1.95(11) \\
\hline
\end{tabular}
%\end{minipage}
\end{table*}

\section{Time-series spectroscopy}
We performed time-series spectroscopy on the night starting on 7 Jan 2010. We used ALFOSC at the Nordic Optical Telescope (NOT) to obtain 472 low resolution spectra. We used grism\#16 and, for all but the first few exploratory spectra, we used a 1.3\,-\,arcsec wide vertical slit. Our dataset spans a total of 6 hours and 40 minutes.

We used a small window and 2$\times$2 binning in the spatial direction to minimize readout-overheads; the window allowed for 28\,arcseconds of sky on either side of the stellar spectrum. To avoid under-sampling and to not smear the dominant variability too much, we set the exposure time to 30\,sec, achieving a 42\,-\,sec cycle time. In the course of our run we obtained spectra of a helium arc lamp to calibrate the wavelength.

\begin{figure*}
\includegraphics[width=170mm]{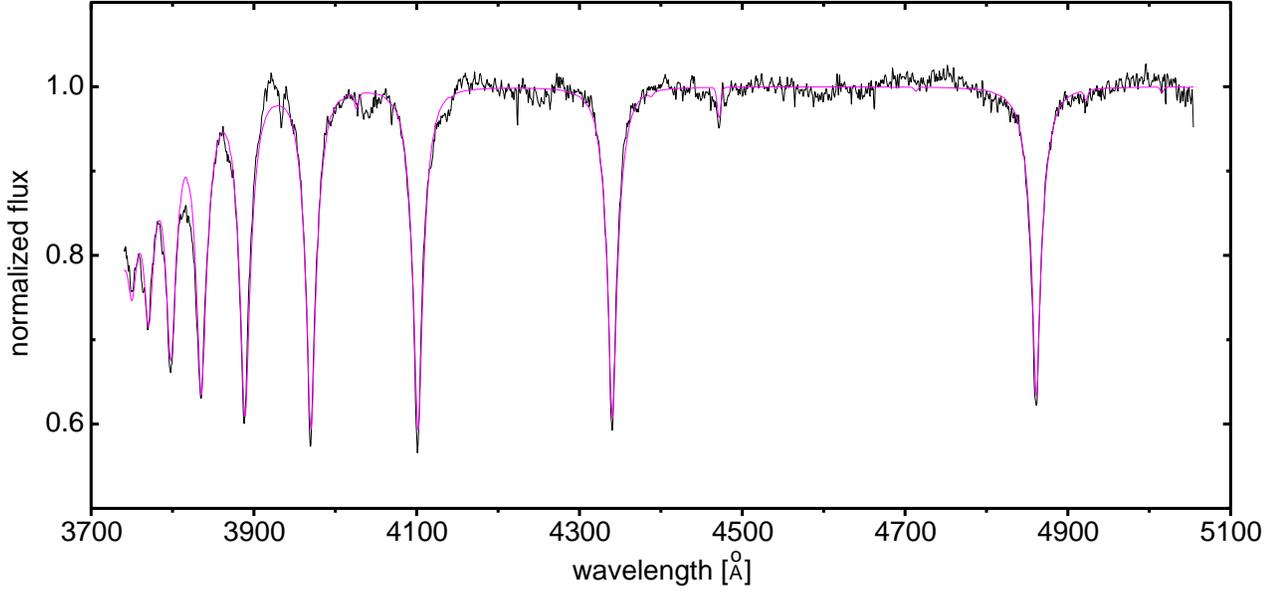}
\caption{Average spectrum of J08069+1527 obtained at NOT and calculated from 472 single low resolution spectra. Magenta line represents a model spectrum used for the final normalization.}
\label{averspec}
\end{figure*}

Unfortunately, since the brightness of this star is about 14\,mag in the B filter only, we could not achieve very high S/N ratio. For single spectra we had typical S/N$\sim$15 in the 440-480 nm region, while we reached S/N$\sim$150 in the average spectrum shown in Fig.\ref{averspec}.

\begin{figure*}
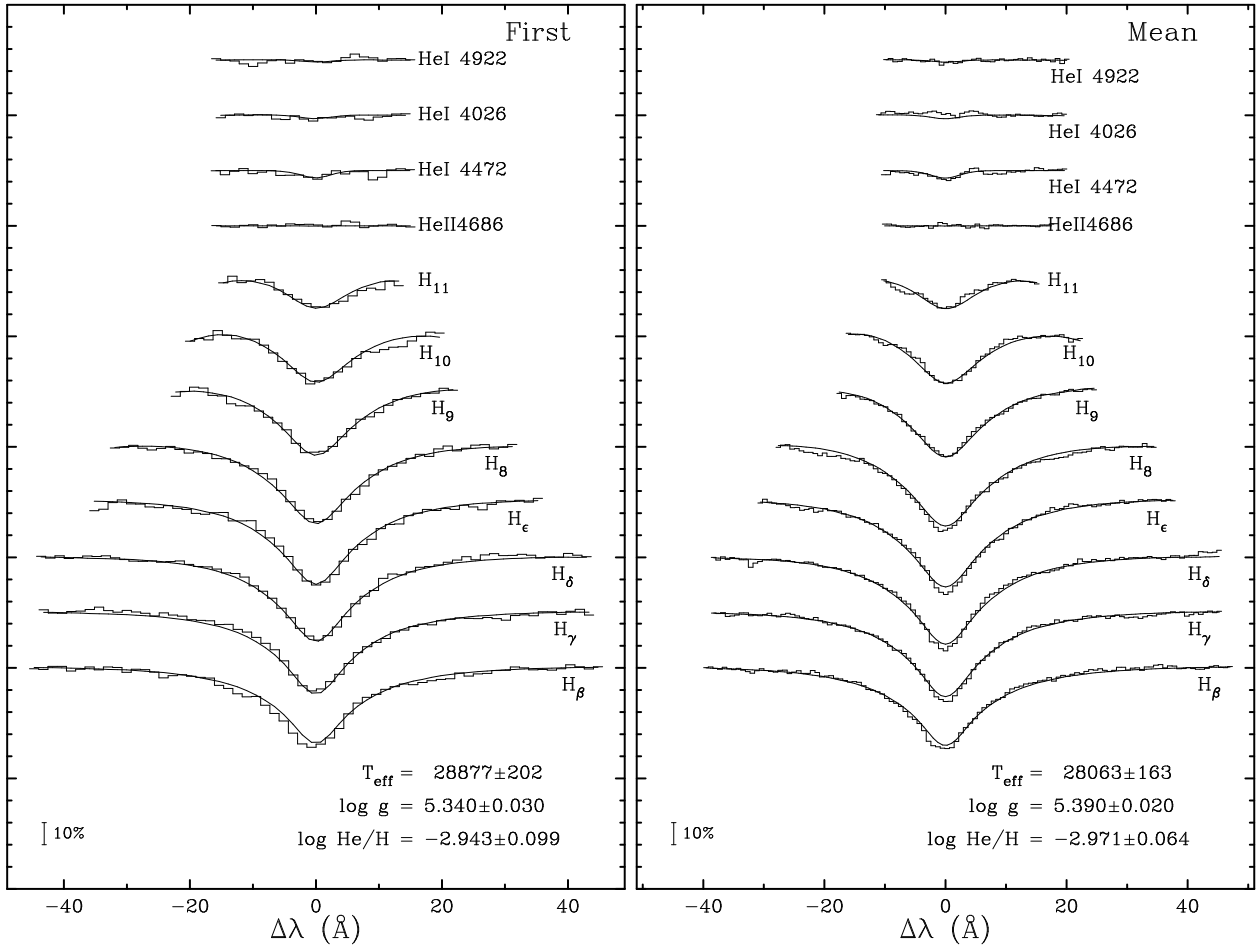

\includegraphics[width=83mm]{fig7a.eps}
\includegraphics[width=83mm]{fig7b.eps}
\caption{Fits to the line profiles of hydrogen and helium lines in the spectrum taken in January 2009 (on the left) and the mean spectrum of recent time-series spectroscopy (on the right) with RV fixed at 335\,km/s. Cited errors are the formal fitting errors between the observed and model spectra.}
\label{spfit}
\end{figure*}

We used standard {\em IRAF} procedures to reduce and calibrate the spectra, leading to 472 extracted spectra. We normalized the average spectrum to unity, using a low-order polynomial fit. The resulting spectrum was then fitted with a model spectrum, and the resulting model spectrum was used for guidance in the final normalization of the average spectrum. The stellar parameters we derived from the average spectrum ($T_{\rm eff}$, $\log g$ and $\log He$ in Fig.\ref{spfit}) were consistent for different normalization/rectification methods that either employ straight-line fits around each Balmer line or a multi-order polynomial fit that rectifies the complete spectrum at once. Then all individually extracted spectra were rectified using the fit that was needed to normalize the average spectrum. Finally the individual spectra were scaled to get their continuum at unity, using a simple parabolic fit in the regions with no absorption lines.

We derived $T_{\rm eff}$\,=\,28,063$\pm$163\,K, $\log g$\,=\,5.39$\pm$0.02 and $\log He/H$\,=\,-2.971$\pm$0.064\,dex. These values are different than those derived from spectrum taken in January 2009 so we adopt an average value for all three parameters: $T_{\rm eff}$\,=\,28,500$\pm$400\,K, $\log g$\,=\,5.37$\pm$0.05\,dex and $\log He/H$\,=\,-2.96$\pm$0.10. The reader should be aware, though, that systematic effects from the model grid can give shifts on the order of 2000\,K and 0.2\,dex in $T_{\rm eff}$ and $\log g$, respectively.

\begin{figure}
\includegraphics[width=83mm]{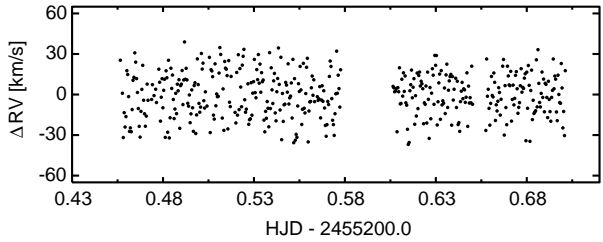}
\caption{Radial velocity displacement in function of time derived from time-series spectroscopy obtained at the NOT.}
\label{rv}
\end{figure}

\begin{figure}
\includegraphics[width=83mm]{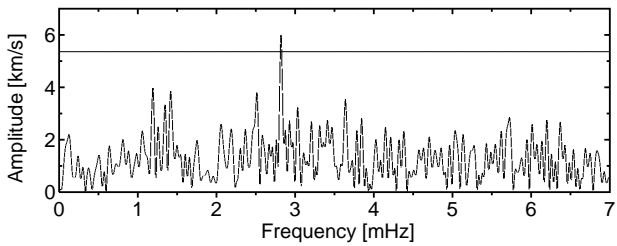}
\caption{Amplitude spectrum calculated from spectroscopic data taken at NOT.}
\label{not-ft}
\end{figure}

\begin{figure}
\includegraphics[width=83mm]{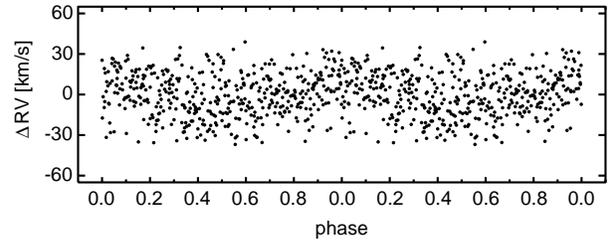}
\caption{Radial velocity displacement folded with the period derived from our analysis. The phase goes over two cycles to better see detected periodicity.}
\label{rv-ph}
\end{figure}

To look for radial velocity (RV) displacements we used the {\em IRAF} cross-correlation routine {\em FXCOR}, finding the radial-velocity shift of each individual spectrum with respect to that of the average spectrum. We included only the wavelength regions around 7 Balmer lines (H{\it $\beta$}-H{\it 10}) in the cross-correlation. We fitted a Gaussian function to find the center of the cross-correlation function, which we adopted as a measure of the radial-velocity shift. As a result we obtained Heliocentric Julian Date along with RV shifts and their errors. We present the radial velocity curve in Fig.\ref{rv}. This result was subject to Fourier analysis in order to extract any periodicity from the data. The amplitude spectrum is shown in Fig.\ref{not-ft}. By means of non linear least square fitting, we detected only one peak which satisfies S/N$>$4. Its frequency, amplitude and phase with errors are: 2.8202(42)\,mHz, 6.1(1.0)\,km/s and 5.69(17)\,rad, respectively. The phase is given at HJD\,=\,2455204.577483. We present the radial velocity curve folded with the period obtained in our analysis in Fig.\ref{rv-ph}. The periodicity detected in a Fourier space barely meet our significance level as S/N\,=\,4.55.
%We did not apply a heliocentric velocity shift to the spectra; the difference in heliocentric velocity between the first and the last of  our spectra is only 0.7\,km/s .

%We performed time-series spectroscopy on 7 Jan 2010. We used EQUIPMENT to obtain low resolution spectra. To avoid under sampling and to not smear the dominant variability too much, we set the exposure time to 30\,sec. Unfortunately, since brightness of this star is about 14\,mag in a B filter only, we could not achieve very high S/N ratio. In a single spectrum it is only 10 within 440-480\,nm and 23 in the average spectrum shown in Fig.\ref{averspec}. We used IRAF to calibrate and analyze our spectra. All single spectra were bias corrected and then flat fielded using halogen lamp. In the course of our run we obtained spectra of helium arc lamp to get spectra wavelength calibrated. Before extracting radial velocity displacements we normalized all spectra using the continuum level.

%To look for radial velocity (RV) displacements we used IRAF tasks on individual spectrum. We used the average spectrum as a template and fitted Gaussian function to the strongest Balmer lines. As a result we obtained Heliocentric Julian Date along with RV shifts and their errors. We present the radial velocity curve in Fig.\ref{rv}. This result was subject to Fourier analysis in order to extract any periodicity existed in the data. The amplitude spectrum calculated from these data is shown in Fig.\ref{not-ft}. By means of non linear least square fitting we detected only one peak which satisfy S/N$>$4 rule. Its frequency, amplitude and phase with errors are: 2.8178(48)\,mHz, 4.93(96)\,km/s and 4.3(2)\,rad, respectively. The phase is given at HJD\,=\,2455204.5728157. We present the radial velocity curve folded with the period obtained in our analysis in Fig.\ref{rv-ph}. Periodicity detected in a Fourier space is barely seen in this figure (S/N\,=\,4.16). Normally we could compare phases of the mode detected in both spectroscopy and photometry to derive radial displacement of the star caused by stellar oscillations. In this case big errors in the result of spectroscopic data does not allow us to do so reliably.

\section{Summary and conclusion}
In this paper we presented our discovery of a new hybrid pulsating subdwarf B star, J08069+1527. Effective temperature and surface gravity place this object in the instability strip among other hybrid subdwarfs. From our analysis of photometric data we detected four peaks, including two in the low frequency domain, confirming our prediction. As the star is relatively faint, spectroscopic data is not sufficient to detect more than just the dominant mode we found in the discovery data. However, it proved that time-series spectroscopy at 2\,m class telescope can support our analysis with radial velocity shifts down to 14\,mag hot subdwarf stars, particularly when data are taken over several nights.

%As it turned out from our analysis, physical parameters are not the only similarities to other hybrid sdBV stars. Frequency of the dominant mode is also similar to those mentioned by \cite{baran09} in Fig.\,9, particularly to Balloon\,090100001. It might be the case that these stars are in very similar evolutionary status and/or have very similar stellar structure. If so, we could use the result obtained by \cite{baran08} to constrain the degree of the dominant mode to $\ell$\,=\,0. This might be supported by the fact that amplitude ratio of the dominant mode in light and radial velocity curves found in J08069+1527 is very similar to the one in Balloon\,090100001.

As it turns out,  physical parameters are not the only similarities J08069+1527 has to other hybrid sdBV stars. As shown in \citet[][in Fig.\,9]{baran10}, some hybrids share a common frequency for their dominant variation and J08069+1527 also has this characteristic. It might be that these stars are in very similar evolutionary state and/or have very similar stellar structure. If so, we could use the result obtained by \cite{baran08} to constrain the degree of the dominant mode to $\ell$\,=\,0. Balloon\,090100001 has photometrically constrained pulsation modes using multiplets \citep{baran09}. The dominant periodicity in Balloon\,090100001, associated with an $\ell =0$ mode, has a photometric amplitude in a B filter of 53\,mma and a radial velocity (RV) amplitude of 19.2\,km/s. The amplitude ratio of RV/B is 0.362. The dominant periodicity in J08069+1527 has an amplitude of 27\,mma in a B(Geneva) filter and RV\,=\,6.1km/s which results in a ratio of 0.226. The difference in ratio is substantial, however, the amplitude spectrum of J08069+1527 might be unresolved and the amplitude of the dominant mode affected by poor resolution. More data of J08069+1527 taken over several days may help to better compare results on these two interesting stars.

%something like, "since BA09 and Jay have virtually the same Teff and logg, and the pulsations have the same period, then assuming XX is radial, in BA09, the photometric amplitude is XX and the RV amplitude is XX. Therefore, since the photometric amplitude of Jay is XX, the RV amplitude should be XX, and it actually is XX (agree or not)

\section*{Acknowledgments}

AB gratefully appreciates funding from Polish Ministry of Science and Higher Education under project No. 554/MOB/2009/0. JTG was supported by the Missouri Space Grant Consortium, funded by NASA. R.H.{\O} acknowledges funding from the European Research Council under the European Community's Seventh Framework Programme (FP7/2007--2013)/ERC grant agreement N$^{\underline{\mathrm o}}$\,227224 ({\sc prosperity}), as well as from the Research Council of K.U.Leuven grant agreement GOA/2008/04. Based on observations made with the Nordic Optical Telescope, operated on the island of La Palma jointly by Denmark, Finland, Iceland, Norway, and Sweden, in the Spanish Observatorio del Roque de los Muchachos of the Instituto de Astrofisica de Canarias.

\label{lastpage}

\end{document}